\documentclass[sigconf]{acmart}

\AtBeginDocument{%
	\providecommand\BibTeX{{%
			\normalfont B\kern-0.5em{\scshape i\kern-0.25em b}\kern-0.8em\TeX}}}

\setcopyright{acmcopyright}
\copyrightyear{2019}
\acmYear{2019}
\acmDOI{10.1145/1122445.1122456}

\usepackage{multirow}
\usepackage{soul}
\usepackage{mathtools}
\usepackage{amsmath,amssymb,amsfonts, amsthm}
\usepackage{algorithmic}
\usepackage{graphicx}
\usepackage{float}
\usepackage{subfig}
\usepackage{textcomp}
\usepackage{xcolor}
\usepackage{placeins}
\usepackage{hhline}
\usepackage{tabularx}
\usepackage{enumitem}
\usepackage{setspace}
\usepackage{booktabs}
\usepackage{adjustbox}
\usepackage[normalem]{ulem}

\theoremstyle{definition}
\newtheorem{problem}{\textbf{Problem}}

\newtheorem{dfn}{Definition}
\let\sigproof\proof\let\proof\relax
\let\sigendproof\endproof\let\endproof\relax
\let\proof\sigproof
\let\endproof\sigendproof

\DeclareMathOperator{\E}{\mathcal{E}}
\DeclareMathOperator{\MLP}{MLP}
\DeclareMathOperator{\Norm}{Norm}
\DeclareMathOperator{\RF}{RF}

\begin{document}
	
	\title{SocialGrid: A TCN-enhanced Method for Online Discussion Forecasting}
	
	\author{Chen Ling}
	\affiliation{%
		\institution{University of Delaware}
		\streetaddress{Department of Information and Computer Science}
		\city{Newark}
		\state{DE}
	}
	\email{lingchen@udel.edu}

	\author{Ruiqi Wang}
	\affiliation{%
		\institution{University of Delaware}
		\streetaddress{Department of Information and Computer Science}
		\city{Newark}
		\state{DE}}
	\email{wangrq@udel.edu}

	\author{Guangmo Tong}
	\affiliation{%
		\institution{University of Delaware}
		\streetaddress{Department of Information and Computer Science}
		\city{Newark}
		\state{DE}}
	\email{amotong@udel.edu}

	
	\begin{abstract}
		As a means of modern communication tools, online discussion forums have become an increasingly popular platform that allows asynchronous online interactions. People share thoughts and opinions through posting threads and replies, which form a unique communication structure between main threads and associated replies. It is significant to understand the information diffusion pattern under such a communication structure, where an essential task is to predict the arrival time of future events. In this work, we proposed a novel yet simple framework, called SocialGrid, for modeling events in online discussing forms. Our framework first transforms the entire event space into a grid representation by grouping successive evens in one time interval of a particular length. Based on the nature of the grid, we leverage the Temporal Convolution Network to learn the dynamics at the grid level. Varying the temporal scope of an individual grid, the learned grid model can be used to predict the arrival time of future events at different granularities. Leveraging the Reddit data, we validate the proposed method through experiments on a series of applications. Extensive experiments and a real-world application. Results have shown that our framework excels at various cascade prediction tasks comparing with other approaches. 
	\end{abstract}
	
	\keywords{time-series forecasting, online discussion forum, temporal convolution network}
	
	\maketitle
	
	\section{Introduction}
	According to the report from Statista\footnote{www.statista.com/statistics/278414/number-of-worldwide-social-network-users/}, there are nearly $3$ billion active users in 2019 that are using online social networks as their tools of daily communication and social interaction. In these years, the growing number of users spurs a large volume of research in the field of online social networks. Research has been devoted to understanding the diffusion dynamics in popular social networks (e.g., Twitter and Facebook), and the existing models \cite{centola2007cascade, cheng2014can} have successfully characterized the information cascades in network-based social media, which derives plenty of real-world applications, such as misinformation containment \cite{nguyen2012containment}, political campaigning \cite{vergeer2013online}, and product marketing \cite{palmer2009experiential}.

    Among the current social platforms, online discussion forums (ODFs) have grown as an important branch and drawn tremendous attention in recent studies \cite{thomas2002learning, gao2013designing, wang2011learning, goggins2016building}. The study of information cascades in ODFs has enabled important applications such as thread recommendation in educational ODFs \cite{lan2018personalized}, latent community discovery \cite{liu2019latent}, and user activity prediction \cite{romero2013predicting}. ODFs are distinct from other social networks because of their particular information sharing pattern: people raise a discussion or a question through posting a thread, and users can respond to such a thread by posting replies, which forms a unique \textit{thread-reply} structure. In this paper, we target on predicting the arrival time of future events in an ODF and explore how such thread-reply structure can be introduced in the design of machine learning methods. 
    
    \textbf{Challenges.} There are a few challenges in predicting the arrival time of future events in ODFs:
    
    \begin{figure*}[!t]
		\subfloat[A Typical ODF]{\label{fig:1a}
			\includegraphics[width=0.5\textwidth]{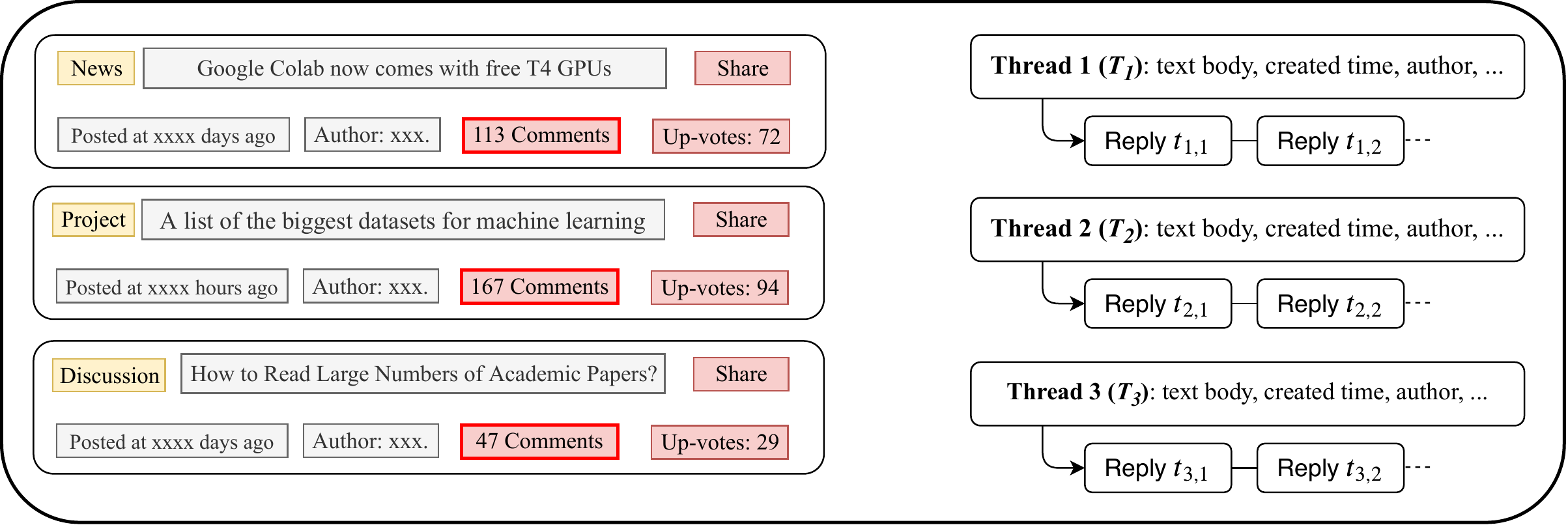}}
		\subfloat[Multi-sequence View of an ODF]{\label{fig:1b}
			\hspace{3mm}\includegraphics[width=0.23\textwidth]{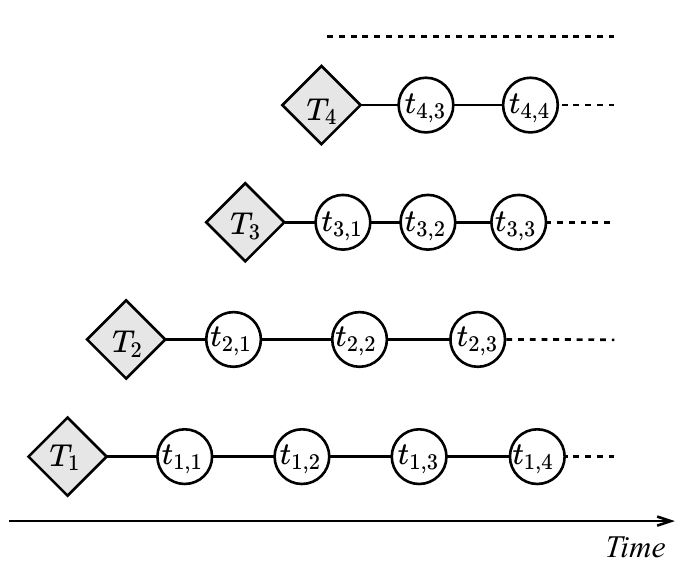}}
		\subfloat[An ODF in the Grid]{\label{fig:1c}
			\hspace{3mm}\includegraphics[width=0.21\textwidth]{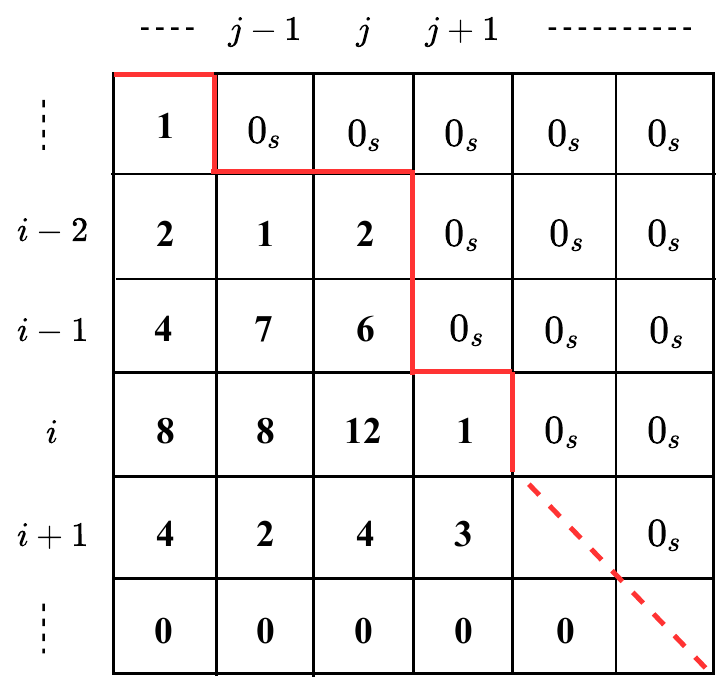}}
		\caption{The transformation pipeline from the raw online social forum to the proposed grid representation. In Fig. \ref{fig:1c}, suppose the length $d$ of the time duration between any two consecutive rows is $5$ minutes, for the $j$-th cascade, $\mathcal{G}_{i, j}$ represents that there are $12$ events observed in the $i$-th $5$ minutes. Besides, the mark $0_s$ on $\mathcal{G}_{i-1, j+1}$  denotes the main thread $T_{j+1}$ has not arrived yet in the $(i-1)$-th time interval. }
		\label{fig:distribution}
	\end{figure*}
    
    \begin{itemize}
	\item \textbf{Complex Hypothesis Space.} Directly predicting the arrival time of each future event requires us to search a complex hypothesis space, making the learning problem theoretically challenging. This is further execrated by the fact that event streams are affected by various exogenous factors that cannot be inferred from training data. 

	\item \textbf{Limited Features.} Employing extra features (e.g., textual content and user profile) can often facilitate a supervised learning task, but their usage is prohibitive when we have to adaptively predict a sequence of future events. For example, if the texture content were required as an input to predict future threads, the adaptive prediction can be performed only if we could predict the future texture content as well. Similarly, we cannot utilize user profile unless we could also predict who is person to post the next thread. In this paper, we seek to design a learning method that only utilizes temporal-related feature.
	

	\item \textbf{Cold Start Problem.} Different from the multi-sequence modeling with a fixed dimension, new threads can be posted dynamically in an ODF. Predicting the first several replies in a newly posted main thread is challenging as very limited information on the current thread is available, which suggests that it is necessary to consider the correlation between different threads.
	
    \end{itemize}
    
    In coping with these challenges, we propose a novel framework called \textit{SocialGrid} for modeling the information diffusion dynamics and predicting the arrival of future events in ODFs. The proposed framework enjoys a few merits:
    
\begin{itemize}
    \item \textbf{Unified Representation.} We propose a simple yet robust method to represent the thread-reply cascades by a grid structure, where each cell denotes the number of events within a short interval. We call this representation the \textit{Grid}. The \textit{Grid} can approximate the original hypothesis space arbitrary well, controlled by the length of the time interval. With such a transformation, we turn to learn the dynamics over cells, which is more tractable than predicting the arrival time of each individual event. 
    
    \item \textbf{Learning Correlations with Temporal Features.} The grid representation offers two advantages. First, the nature of grid allows it to be effectively learned through convolutional neural networks. In particular, we extend the one-dimension Temporal Convolutional Network (TCN) to a multi-dimension level and design two models for predicting the arrival time of threads and replies, respectively. Second, by sliding the kernel over the entire input, the proposed model can capture the correlations between different threads, and therefore addresses the cold start problem implicitly.
    
    
    
    
    \item \textbf{High Practical Performance.} We validate our model by  experiments on real-world datasets. Comparing with the state-of-the-art approaches such as RNN, ARIMA, and Temporal Point Process, we have observed that the performance of our model is better than others by an evident margin in two applications: predicting arrival times and identifying breakout cascades. Our code and dataset are publicly available.\footnote{https://github.com/lingchen0331/SocialGrid}
\end{itemize}
    

        
        
    
    \textbf{Road-map.} Sec. \ref{sec: 2} introduces the problem setting and the grid transformation of an ODF. In Sec. \ref{sec: 3}, we present the learning strategy. Sec. \ref{sec: exp} describes the experiment procedures and shows the result. We survey the related works and conclude the paper in Secs. \ref{sec: related} and \ref{sec: con}, respectively.  
    \vfill\null
	

    
    \section{Problem Setting and Grid Transformation}\label{sec: 2}
    An ODF is composed of two types of event streams, \textit{thread stream} $\{T_i\}$ and \textit{reply stream} $\{t_{i, j}\}$, where each thread $T_i$ is posted as a new web content to which users can respond through the associated replies $\{t_{i, 1}, t_{i, 2}, ..., t_{i, j}, ...\}$, as shown in Fig. \ref{fig:1a}. We slightly abuse the notation and use $T_i \in \mathbb{R}^+$ and $t_{i,j} \in \mathbb{R}^+$ to also denote the arrive times of the events. For the multivariate thread-reply cascades, we in this paper study the problem of forecasting the future event times. Formally, it is stated as  
    \begin{problem}[\textbf{Time-series Forecasting}] 
    \label{prob: 1}
    Given the arrival times $\E_{\leq t}$ of the events before a time point $t\in \mathbb{R}^+$, we aim to learn the conditional distribution
    \begin{equation}
    \label{eq: prob_1}
    \Pr[\mathcal{E}_{> t}|\mathcal{E}_{\leq t}]
    \end{equation}
    where $\E_{> t}$ denotes the arrival time of the events occurring after $t$.
    \end{problem}
    \noindent
    Directly predicting $\mathcal{E}_{> t}$ conditioned on history $\mathcal{E}_{\leq t}$ is difficult. However, By placing events from each information cascade in the same timeline, as shown in Fig. \ref{fig:1b}, we can transform the whole event space of an ODF as a grid-shaped representation, which we call it the \textit{Grid}. 
    
     
    
    \begin{dfn}[\textbf{Grid}]
    Grid $\mathcal{G}$ is a matrix where each column $j$ denotes an individual thread-reply cascade with main thread $T_j$, and each row $i$ represents a time interval of length $d \in \mathbb{R}^+$. The Grid consists of two types of elements. For each $\mathcal{G}_{i,j}$, if the main thread $T_j$ has already arrived in or before the $i$-th time interval, then it denotes the total number of events of this $j$-th cascade arrived in the $i$-th time interval. Otherwise, it is denoted as $0_s$. A binary mask matrix $\mathcal{M}$ with the same shape as $\mathcal{G}$ is employed to record the position of all $0_s$: 
    \begin{equation}\label{eq: mask}
       \mathcal{M}_{i, j} = 
        \begin{cases}
            1 & \text{if $\mathcal{G}_{i, j} = 0_s$,} \\
            0 & \text{otherwise.}
        \end{cases}
    \end{equation}
    \end{dfn}
    \noindent
    An illustration of the Grid is given in Fig. \ref{fig:1c}. In this work, we use the Grid as an alternative but simple way to represent an ODF, and the Problem \ref{prob: 1} can thus be transformed accordingly.  
    \begin{problem}[\textbf{Time-series Forecasting in the Grid}] \label{prob: 2}
    Given the Grid $\mathcal{G}$, we aim to learn the distribution of elements $\mathcal{G}_{\ge i, j}$ conditioned on $\mathcal{G}_{< i, j}$:
    \begin{equation} \label{eq: prob_2}
         \Pr[\mathcal{G}_{\ge i, j}|\mathcal{G}_{<i, j}],
    \end{equation}
    where $\mathcal{G}_{\ge i, j}$ denotes the elements that are currently at or beyond the $i$-th time interval, and $\mathcal{G}_{<i, j}$ denotes all the historical elements that have arrived before the $i$-th time interval.
    \end{problem}
   
    Avoid searching the complex hypothesis space, the conditional probability in Eq. \ref{eq: prob_2} can be estimated by conducting convolution-related operations on the Grid. Through learning the Grid, we can forecast the number of events in each of the following time intervals. Also, we can precisely predict the arrival time of the upcoming events if we shrink the time interval length $d$ in the Grid. In the next section, we discuss the learning strategy and design two models of predicting the arrival time of different types of events in the Grid, respectively.
 
    \section{Learning Strategy}\label{sec: 3}
    \subsection{Preliminaries}
    TCN \cite{bai2018empirical} is introduced as a variant of convolution architecture that is specialized in the area of one-dimension temporal sequence modeling. Particularly, TCN features two essential techniques - causal convolution and dilated convolution.

    \textbf{Causal Convolution.} For a one-dimension (1-D) input sequence $\mathbf{x} \in \mathbb{R}^{n}$, a traditional 1-D CNN layer returns a output $\mathbf{y} \in \mathbb{R}^{n-q+1}$, where $q \in \mathbb{Z}^{+}$ is the size of the convolution filter. Particularly for sequence modeling, causal convolution adds padding to keep the output sequence having the same size as the input sequence. Let $\Gamma^{l}(i)$ be the hidden representation of the element at the time step $i$ of the $l$-th layer, and $\mathbf{f} \in \mathbb{R}^{q}$ is the causal convolution filter with size $q$. A causal convolution at the $l$-th layer can be written as:
    \begin{equation}\label{eq: 3}
        \Gamma^{l}(i) = \sum^{q}_{j=1} \mathbf{f}_j \cdot \Gamma^{l-1}(i-j), l\ge 1,
    \end{equation}
    where only the elements arrived before the $i$-th time step can be convoluted with the filter. Therefore, causal convolution is able to ensure the temporal causality in the temporal sequence modeling. 
    
    \textbf{Dilated Convolution.} The traditional 1-D CNN is also restricted in its limited receptive field. Simply stacking of convolution layers can increase the receptive field, but the network complexity also increases. TCN employs the dilated convolution to enlarge the receptive field when stacking with more convolution layers. The combination of the dilated convolution and the causal convolution is written as: 
    \begin{equation*}
        \Gamma^{l}(i) = \sum^{k}_{j=1} \mathbf{f}_j \cdot \Gamma^{l-1}(i-j\cdot \tau), l\ge 1,
    \end{equation*}
    where $\mathbf{f}\in \mathbb{R}^q$ is the convolution filter with a dilation rate $\tau \in \mathbb{Z}^+$. With several such convolution layers, the TCN model can access a sufficiently large receptive field. In addition, to further reduce the risk of network degeneration, the 1-D TCN also utilizes residual connection \cite{he2016deep} as well as ourput normalization between every two dilated causal convolution layers.
    
    The traditional TCN is designed for modeling $1$-D sequence only, and it cannot be used in our scenario. Therefore, we extend the origin $1$-D TCN to multi-dimension ($2$-D) level so that the time series prediction in the Grid can be estimated through convolution operations while remaining temporal causality.
    
    
	\subsection{$2$-D TCN}
	In this work, we leverage a stack of $2$-D temporal convolution layers to learn the direct mapping from $\mathcal{G}_{< i, j}$ to $\mathcal{G}_{\ge i, j}$. Each $2$-D temporal convolution layer follows a $2$-D dilated causal convolution and a residual connection, which we introduce them respectively.
	
	\subsubsection{$2$-D Causal Convolution.} 
	To learn the latent correlation in the Grid, we extend the architecture of causal convolution from the 1-D TCN. Similar with the causal convolution in $1$-D TCN, the prediction of $\mathcal{G}_{\ge i, j}$ only depends on the previous information $\mathcal{G}_{< i, j}$ before the $i$-th time interval. Thus, a formal expression of 2-D causal convolution at $l$-th layer can be written as:
    \begin{equation*}
        \Lambda^{l}(i, j) = \sum^{K}_{k_1 = 1}\sum^{K}_{k_2 = 1} \mathbf{f}_{k_1, k_2} \cdot \Lambda^{l-1}(i-k_1, j-k_2), l\ge 1,
    \end{equation*}
    where the $\Lambda^{l}(i, j)$ represents the hidden representation of the prediction $\mathcal{G}_{i, j}$ at the $l$-th layer, and $\mathbf{f} \in \mathbb{R}^{K\times K}$ is a square-shaped convolution filter with size $K \in \mathbb{Z}^+$. Note that the $\Lambda^{0}(i, j)$ represents the element $\mathcal{G}_{i-1, j}$, and this formulation ensures that the prediction of $\mathcal{G}_{i, j}$ only depends on the previous elements $\mathcal{G}_{<i, \le j}$. 
	
	\subsubsection{$2$-D Dilated Convolution.} 
	Similar with 1-D TCN, we apply the dilated convolution to efficiently increase the receptive field when stacking multiple convolution layers. We give the formal expression of the causal convolution with a dilation rate $\tau$ in the $l$-th layer as follows.
	\begin{equation} \label{eq: 7}
        \Lambda^{l}(i, j) = \sum^{K}_{k_1=1}\sum^{K}_{k_2=1} \mathbf{f}_{k_1, k_2} \cdot \Lambda^{l-1}(i-k_1\cdot \tau, j-k_2\cdot \tau), l\ge 1,
    \end{equation}
    where the dilation rate $\tau \in \mathbb{Z}^+$ increases exponentially with the number of stacked $2$-D dilated convolution layers (i.e., $\tau = O(2^l)$ at the $l$-th layer). The dilation rate ensures to achieve an exponentially large receptive field in a deep network structure. If $\tau = 1$, a dilated convolution reduces to a regular causal convolution as in Eq. \ref{eq: 7}. Normally, the exponential growth of the dilation rate is added in each of the causal convolution layers to get a larger receptive field over layers \cite{bai2018empirical}. With the help of the dilated convolution, we have two ways to increase the receptive field: either set a large filter size $K$ or stack with more convolution layers. We give the recursive formula of calculating the receptive field $\RF_l$ at the $l$-th layer of the dilated causal convolution:
    \begin{figure}[tbp]
		\centerline{\includegraphics[width=0.45\textwidth]{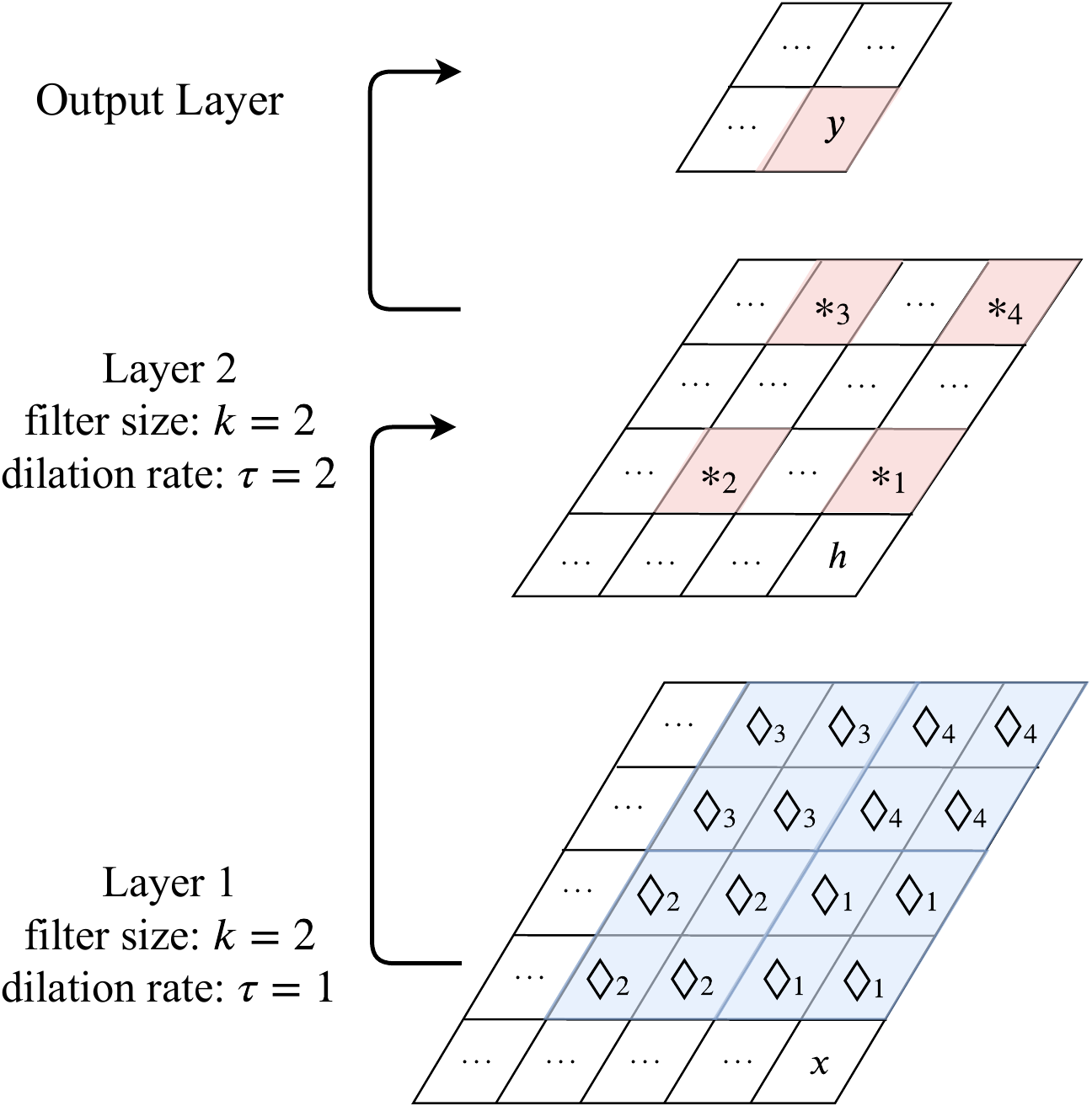}}
		\caption{The visualization of the 2-D dilated causal convolution in SocialGrid: suppose the the element $x$ in the first layer corresponds to the element $y$ in the output layer. As shown in the figure, the element $y$ is decided by $*_i, i \in [1, 2, 3, 4]$ at Layer $2$, and each of the $*_i$ element is decided by the $\Diamond$ element with the same index number. By stacking with more dilated 2-D causal convolution layers, the output element $y$ will be decided by a larger receptive field.}
		\label{fig: 2}
	\end{figure}
	
	\begin{equation*}
       \RF_l = 
       \begin{cases}
       \Big((\RF_{l-1}-1)+ K +(K-1)\cdot(\tau -1)\Big)^2 & l \ge 2\\
       K^2 & l = 1
        \end{cases}.
    \end{equation*}  
	
	In summary, a dilated 2-D causal convolution layer performs a discrete-time dilated causal convolution operation on the input $\mathcal{G}_{>i, j}$ and add a learnable bias term to its output $\mathcal{G}_{\le i, j}$. To better understand the whole process of the 2-D dilated causal convolution, we present the illustration in Fig. \ref{fig: 2}.
    
	\subsubsection{Temporal Convolution Block.}
	Each $2$-D dilated causal convolution layer follows a normalization layer and a non-linear activation function to reduce the internal covariate shift. We present the following equation of the final output from the $l$-th layer.
	\begin{equation*}
	    \Lambda^{*} = \sigma(\Norm(\Lambda^{l})),
	\end{equation*}
	where the $\Lambda^{*}$ is the output from the $l$-th convolution layer, $\Norm$ represents the normalization layer, and $\sigma$ denotes the activation function. The output $\Lambda^{*}$ is further served as the input of the $l+1$-th convolution layer. We refer to the presented architecture as a temporal convolution block.
	
	Additionally, by only controlling the dilation rate $\tau$ and the filter size $K$, one still may not access sufficient historical information without a deep network structure. Therefore, to avoid the potential risk of the network performance degeneration caused by the deep network structure, we add a residual connection \cite{he2016deep} between the temporal convolution blocks. The intuition behind the residual connection is to let the model directly learns the modification from the identity mapping of the last block instead of the full transformation. In our design, the prediction in the output $\Lambda^{*}_{i, j}$ is added element-wisely with the input $\Lambda^{l-1}_{i, j}$ from last $l-1$-th layer. Note that we apply an additional fully convolution network to $\Lambda^{*}$ to account for discrepant shapes between $\Lambda^{*}$ and $\Lambda^{l-1}$. Finally, the target cell $\mathcal{G}_{i, j}$ can be predicted as:
	\begin{equation*}
		\mathcal{G}_{i, j} = \sigma(\Lambda^*_{i, j} \oplus \Lambda^{l-1}_{i, j}).
	\end{equation*}
	
\begin{figure}[tbp]
		\centerline{\includegraphics[width=0.49\textwidth]{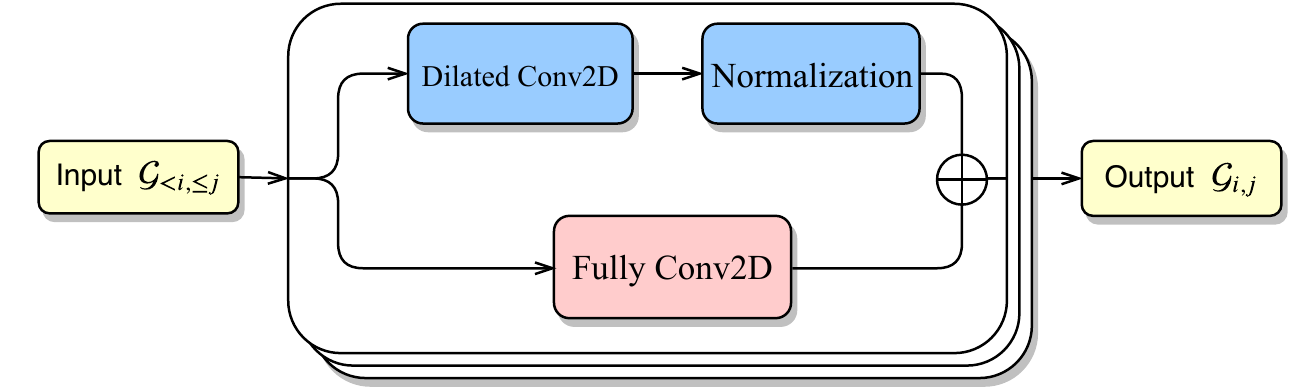}}
		\caption{The Temporal Convolution Block}
		\label{fig: 3}
	\end{figure}

	The whole process is summarized in Fig. \ref{fig: 3}, including the dilated 2-D causal convolution layer, the normalization layer, and the residual connection. Note that the normalization is taken to be Batch Normalization, and we use the Parametric Rectified Linear Unit (PReLU) \cite{he2015delving} as the activation function in all of the temporal convolution block. Based on the design of temporal convolution block, we design two models of predicting the arrival time of main threads and replies, respectively.

    \subsection{Model of Main Threads}
    To predict the arrival time of future main threads, we propose a model that can simulate the number of $0_s$ between two main threads. Formally, our model is designed to learn the following conditional probability:
    \begin{equation*}
        \Pr[O^{j}_{j+1}|\mathcal{G}_{\le i, \le j}],
    \end{equation*}
    where the $O^{j}_{j+1} \in \mathbb{N}$ is the number of $0_s$ between two consecutive main threads $T_{j}$ and $T_{j+1}$. Note that the $O^{j}_{j+1}$ can also be viewed as the number of time intervals between $T_{j}$ and $T_{j+1}$. Therefore, we can estimate the arrival time $T_{j+1}$ of the next main thread by:
    \begin{equation}\label{eq: arrival_time}
        T_{j+1} = T_{j} + (O^{j}_{j+1} \cdot d).
    \end{equation}
    The prediction of $T_{j+1}$ in Eq. \ref{eq: arrival_time} is approximate as directly predicting the arrival time by considering the error in terms of the length $d$. This estimation result would be similar as directly predicting the event arrival time if we shrink the time interval length $d$ in the Grid small enough.
    
	\begin{figure*}[!t]
		\subfloat[Predicting the Arrival Time of Future Main Thread]{\label{fig:4a}
			\hspace{3mm}\includegraphics[width=0.45\textwidth]{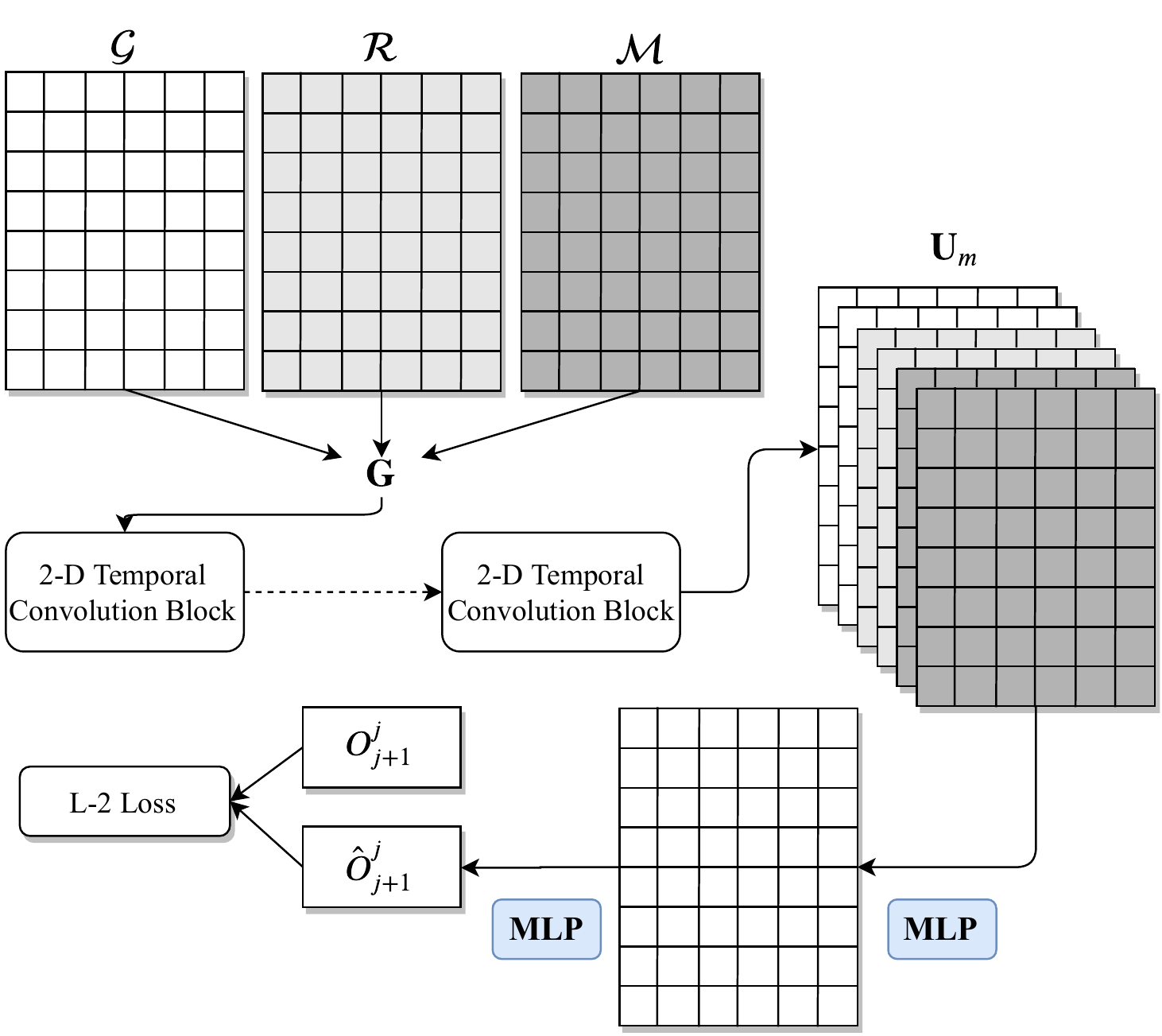}}
		\subfloat[Predicting the Distribution of Future Replies]{\label{fig:4b}
			\hspace{3mm}\includegraphics[width=0.46\textwidth]{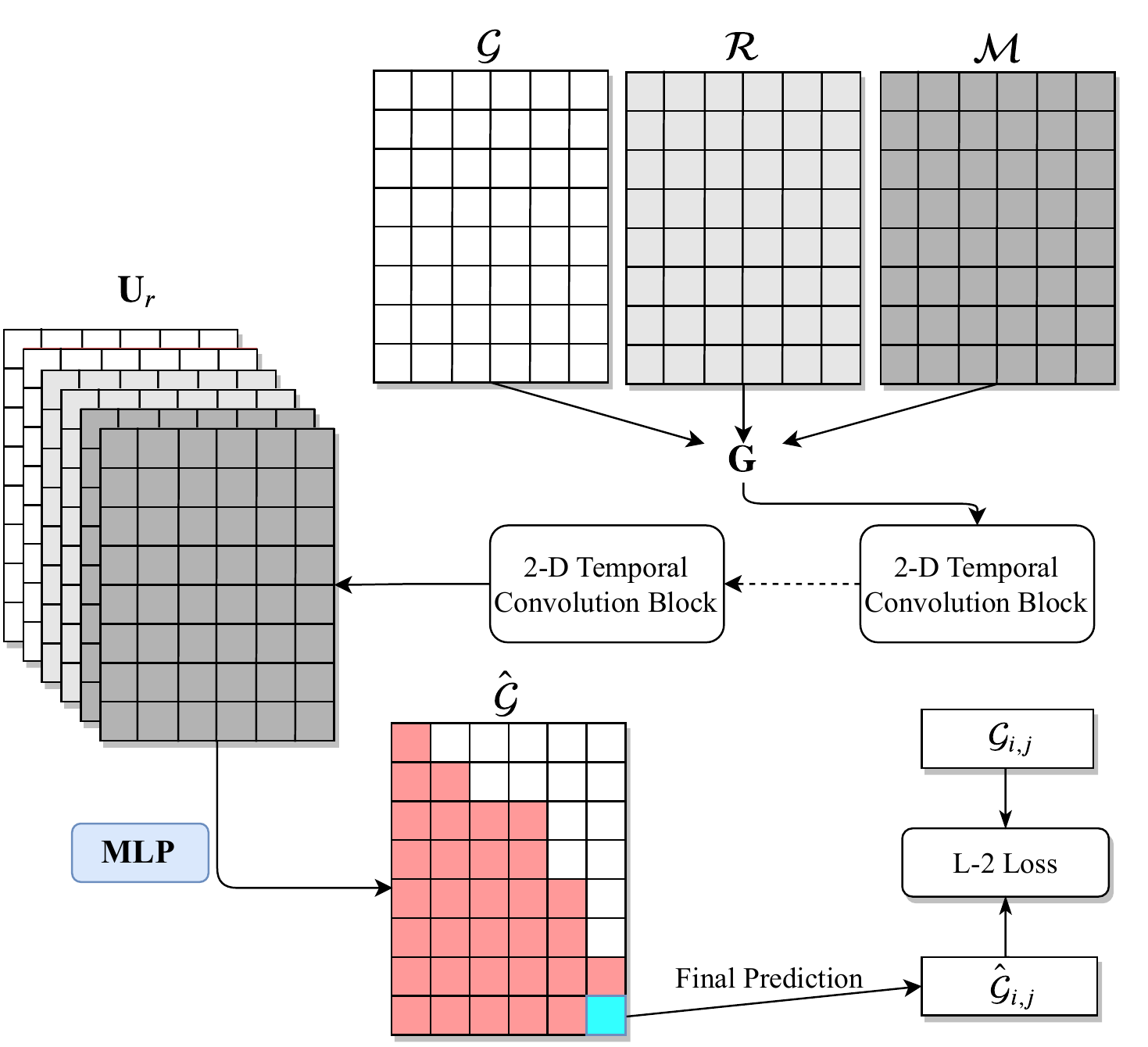}}
		\caption{Each model takes the tensor $\mathbf{G}$ consisting of three feature channels as the input, and $\mathbf{U}_{m}$ and $\mathbf{U}_{r}$ are returned as output tensors, respectively. $\mathbf{U}_{m}$ and $\mathbf{U}_{r}$ are further compressed by $\MLP$ with an activation function to get the final prediction of individual prediction task. For the model of predicting the future thread arrival time, the MSE loss is calculated between the predicted value and the ground truth. For the model of estimating the future reply numbers, we only calculate the MSE loss between the element in the bottom right corner of the matrix and the predict value, as shown in Fig. \ref{fig:4b}.}
	\end{figure*}
    
    \textbf{Input.} To more accurately predict the $O^{j}_{j+1}$, we utilize auxiliary features directly derived from the Grid $\mathcal{G}$. Specifically, feature matrix $\mathcal{R}$ records the number of time intervals (rows) from each cell $\mathcal{G}_{i, j}$ to the time interval of the associated main thread in the Grid $\mathcal{G}$. Note that $\mathcal{R}_{i, j} = 0$ if the corresponding $\mathcal{G}_{i, j} = 0_s$, and we further normalize all the columns in $\mathcal{R}$ to the range of $[0, 1]$ in order to keep the count of each cascade in the same scale. When we predict the $\mathcal{G}_{i, j}$, we utilize the $\mathcal{R}$ to inform our model of the latent influence of its associated main thread. In addition, the mask matrix $\mathcal{M}$ is from Eq. \ref{eq: mask} to remind the model of the location of all $0_s$. Therefore, the input of the 2-D temporal convolution block is composed of three feature matrices, which forms a 3-D tensor $\mathbf{G}$ with three channels $\mathcal{G}$, $\mathcal{R}$, and $\mathcal{M}$.

    \textbf{Model Pipeline.} To get the estimation of $T_{j+1}$, the input of the model a $3$-D tensor $\mathbf{G}$ with feature channels $\mathcal{G}_{\le i, \le j}$, $\mathcal{R}_{\le i, \le j}$, and $\mathcal{M}_{\le i, \le j}$. The tensor $\mathbf{G}$ is convoluted by a certain amount of temporal convolution block to generate an output tensor $\mathbf{U}_{m}$. The output $\mathbf{U}_{m}$ is further compressed by Multi-layer Perceptron ($\MLP$) to obtain the $\hat{O}^{j}_{j+1}$ between the main thread $T_j$ and future thread $T_{j+1}$. According to Eq. \ref{eq: arrival_time} , we can estimate the $T_{j+1}$ in the Grid. The loss is taken to be L-$2$ loss between $\hat{O}^{j}_{j+1}$ and its ground truth. We present the pipeline of this prediction in Fig. \ref{fig:4a}.

	\subsection{Model of Future Replies}
	We also design a model that predicts the number of future replies $\mathcal{G}_{\ge i, j}$ conditioned on history $\mathcal{G}_{< i, j}$ according to Eq. \ref{eq: prob_2}. Similarly, by controlling the length $d$ of the time interval in the Grid, the result of $\mathcal{G}_{\ge i, j}$ is approximate as directly predicting the arrival time of future replies if $d$ is small enough. Or we can also estimate the number of replies of a cascade in a while. 
	
	\textbf{Input.} The Grid $\mathcal{G}$ records the number of events of each thread-reply cascade in consecutive time intervals, and it serves as the primary feature for predicting the number of replies. Same as the model of forecasting the future thread arrival time, we also utilizes the relative time $\mathcal{R}$ and the mask information $\mathcal{M}$ to guide the model better predicting the replies. 

	\textbf{Model Pipeline.} 
	The input tensor $\mathbf{G}$ is convoluted by multiple temporal convolution blocks, and the output tensor $\mathbf{U}_r$ is further compressed by $\MLP$ layers to get a 2-D matrix $\hat{\mathcal{G}}$. Each predicted element $\hat{\mathcal{G}}_{i, j}$ corresponds to the element $\mathcal{G}_{i+1, j}$ in the original Grid. In other words, the $\hat{\mathcal{G}}$ is $\mathcal{G}$ up-shifted by one time interval. We use L-$2$ loss as our loss function. The model pipeline is also illustrated in Fig. \ref{fig:4b}.

	\section{Experiment}\label{sec: exp}
	In this section, we describe the overview of datasets and the experiment settings. We then compare our framework SocialGrid to other baseline approaches in various applications.
	
	\subsection{Experiment Setting} 
	We implement the proposed framework in Tensorflow, and we employ Adam optimizer with a learning rate of $1.0 \times 10^{-3}$ for training. In addition, L-$2$ regularizer ($1.0 \times 10^{-2}$) is applied to all convolution filters in the model of predicting future replies. We execute the grid search strategy to locate the best parameters through validation set: the number of filters is chosen from $[16, 32, 64, 128]$, and the filter size is chosen from $[3, 5, 7, 9]$ in all dilated causal convolution layers. The number of temporal convolution blocks is taken from $[3, 4, 5, 6, 7]$, and the dilation rate $\tau = 2^{i-1}$ at the $i$-th block.
	
	The proposed Grid $\mathcal{G}$ is conceptually an infinite matrix in terms of time, we thus slice the entire Grid $\mathcal{G}$ into segments for training. Each segment contains the information of several thread-reply cascades in consecutive time intervals. We also generate the segments of feature matrices $\mathcal{R}$ and $\mathcal{M}$ accordingly. We apply zero-padding only to the top and left side of the feature matrices to ensure the input-output shape consistency and temporal causality during the convolution operations.

	\textbf{Baselines.} We compare our framework SocialGrid with four types of baselines:
	\begin{itemize}
	    \item Traditional Statistical Model: ARIMA (Auto Regressive Integrated Moving Average) \cite{contreras2003arima} is a class of statistical models in the area of time-series analysis,which explains a given time-series based on its own past values. ARIMA is often chosen to be a fundamental baseline in recent time-series forecasting works. We implement the ARIMA model with the public Python library\footnote{https://www.statsmodels.org}. 
	    
		\item Temporal Point Processes (TPP): A temporal point process \cite{hawkes1971spectra} is a random process whose realizations consist of the isolated event arrival time. TPP models have been viewed as natural choices of modeling diffusion patterns of sequences in recent years, and a number of following works \cite{zhao2015seismic,kobayashi2016tideh} utilize the TPP to model the information propagation pattern in the domain of the social network. We select the most fundamental  TPP model Hawkes process \cite{hawkes1971spectra} and an additional TPP model - NesTPP\footnote{github.com/lingchen0331/NesTPP} that is specified in simultaneously modeling the diffusion of main threads and associated replies. Note that we use public Python library\footnote{pypi.org/project/hawkeslib/} for the Hawkes process implementation.

		\item Recurrent Neural Networks (RNN): Multiple empirical studies \cite{gers2002learning, chung2014empirical} have shown the effectiveness of recurrent architectures in the task of sequence modeling, since the recurrent architectures are designed to learn the long dependencies from the sequence. Many variants of RNN (e.g., LSTM and GRU) have achieved state-of-the-art results in many sequence modeling tasks. In this work, we implement the vanilla GRU and LSTM as the comparison model. Both GRU and LSTM contains $64$ hidden units and $4$ layers of the recurrent structure. 
		
		\item Variants of SocialGrid: We further compare SocialGrid with its variants. Specifically, 1-D TCN \cite{bai2018empirical} is implemented to verify if building the multivariate temporal convolution structure can benefit the prediction accuracy. Additionally, in order to verify the effectiveness of additional features, SocialGrid-S is implemented only with one feature channel $\mathcal{G}$, and SocialGrid-M contains two feature channels of $\mathcal{G}$ and $\mathcal{R}$ but without the feature channel of masks $\mathcal{M}$.
	\end{itemize}
	
	\textbf{Evaluation Metrics.} To measure and evaluate the performance of different methods in predicting the arrival time of main threads and the number of replies in unit time, we adopt the following metrics: Mean Absolute Error (MAE) and Root Mean Square Error (RMSE) to measure the relative error between the ground truth array and the prediction result array.
	
	\subsection{Data}
	Reddit\footnote{www.reddit.com} is one of the largest ODFs, and it provides us a sound dataset to evaluate the performance of our model. We select two thread-reply datasets from different subreddits\footnote{A subreddit is a web forum on Reddit of a particular topic.}, and each dataset contains key attributes of the arrival time of each main thread and associated replies. We select cascades within one subreddit as we assume there are underlying correlations between them that can be inferred by the Grid. A brief overview of the selected dataset is detailed below.
	
	\textbf{NBA Dataset.} This dataset was gathered from one of the most sought-after Reddit subforums - NBA subforum during the 2019 Playoffs, which attracts the most attention in a season. We selected $1,610$ main threads and $71,638$ corresponding replies all related to the famous player - LeBron James, in the time range of one month. We apply main threads and linked replies in first three weeks as the training-validation set, the rest of the data serve as the test set.
	
	\textbf{NFL Dataset.} Similarly, we retrieved all the discussions of the topic \textit{Superbowl 2019} during the week of the NFL final match. This dataset contains $1,895$ main threads and $138,911$ associated replies. We also split threads and corresponding replies arrived in the first $5$ days (around $1,600$ thread-reply cascades) as the training-validation set, and all the thread-reply cascades arrived in the rest two days serve as the test set.

\subsection{Application: Non-adaptive Prediction}\label{sec: non_adaptive}
\subsubsection{Prediction of Main Thread Arrival Time}
	\begin{table}[!t]
\centering
\resizebox{0.46\textwidth}{!}{%
\begin{tabular}{@{}c|c|c|c|c@{}}
\toprule
\multirow{2}{*}{Model} & \multicolumn{2}{c|}{NBA Dataset}        & \multicolumn{2}{c}{NFL Dataset}       \\ \cmidrule(l){2-5} 
                       & \multicolumn{2}{c|}{(MAE/RMSE) in Hour} & \multicolumn{2}{c}{(MAE/RMSE) in Hour} \\ \midrule
ARIMA                  & $1.352\pm 0.21$          & $1.876\pm 0.34$        & $0.591\pm 0.09$         & $0.754\pm 0.07$        \\ \midrule\midrule
\multicolumn{5}{c}{Recurrent Methods}\\ \midrule
GRU                    & $1.241\pm 0.09$          & $1.84\pm 0.09$         & $0.416\pm 0.06$         & $0.613 \pm 0.05$        \\ \midrule
LSTM                   & $1.727\pm 0.1$           & $2.622 \pm 0.19$        & $0.51\pm 0.05$          & $0.784\pm 0.06$        \\ \midrule \midrule
\multicolumn{5}{c}{Temporal Convolution Methods}\\ \midrule
TCN                   & $0.877\pm 0.1$           & $1.132 \pm 0.12$        & $0.348\pm 0.08$          & $0.697\pm 0.12$        \\ \midrule
SocialGrid-S          & $0.594 \pm 0.03$          & $1.112\pm 0.04$        & $0.214\pm 0.03$         & $0.298\pm 0.02$        \\ \midrule
textbf{SocialGrid-M}          & $\mathbf{0.521\pm 0.03}$          & $\mathbf{0.917\pm 0.07}$        & $\mathbf{0.174\pm 0.01}$         & $\mathbf{0.266\pm 0.02}$        \\ \midrule
SocialGrid             & $0.576\pm 0.04$          & $0.924\pm 0.05$        & $0.181\pm 0.03$         & $0.279\pm 0.04$        \\ \bottomrule
\end{tabular}%
}
\caption{Performance comparison of different approaches on both datasets in predicting the main thread arrival time.}
\vspace{-3mm}
\label{tab:main}
\end{table}
	
	    \begin{table}[!t]
\centering
\resizebox{0.46\textwidth}{!}{%
\begin{tabular}{@{}c|c|c|c|c@{}}
\toprule
\multirow{2}{*}{Model} & \multicolumn{2}{c|}{NBA Dataset}        & \multicolumn{2}{c}{NFL Dataset}       \\ \cmidrule(l){2-5} 
                       & \multicolumn{2}{c|}{(MAE/RMSE) in Number} & \multicolumn{2}{c}{(MAE/RMSE) in Number} \\ \midrule
ARIMA                  & $5.053\pm 0.22$          & $6.269\pm 0.24$        & $2.563\pm 0.02$         & $2.902\pm 0.03$        \\ \midrule\midrule
\multicolumn{5}{c}{Recurrent Methods}\\ \midrule
GRU                    & $4.819\pm 0.21$          & $5.903\pm 0.2$         & $2.211\pm 0.07$         & $2.806 \pm 0.08$        \\ \midrule
LSTM                   & $4.629\pm 0.29$           & $5.798 \pm 0.41$        & $2.237\pm 0.12$          & $2.862\pm 0.04$        \\ \midrule \midrule
\multicolumn{5}{c}{Temporal Convolution Methods}\\ \midrule
TCN                   & $2.985\pm 0.14$           & $3.482 \pm 0.31$        & $2.022\pm 0.1$          & $2.578\pm 0.15$        \\ \midrule
SocialGrid-S          & $2.202 \pm 0.09$          & $3.863\pm 0.17$        & $1.628\pm 0.04$         & $2.132\pm 0.06$        \\ \midrule
SocialGrid-M          & $2.063 \pm 0.07$          & $3.829\pm 0.27$        & $1.573\pm 0.03$         & $2.282\pm 0.07$        \\ \midrule
\textbf{SocialGrid}             & $\mathbf{1.971\pm 0.04}$          & $\mathbf{3.813\pm 0.161}$        & $\mathbf{1.524\pm 0.05}$         & $\mathbf{2.164\pm 0.1}$        \\ \bottomrule
\end{tabular}%
}
\caption{Performance comparison of different approaches on both datasets in predicting the number of replies in thread-reply cascades.}
\label{tab:reply}
\vspace{-4mm}
\end{table}

    We first consider the task of predicting the arrival time of the future main threads. For non-SocialGrid approaches, we train them on the actual arrival time of $1,000$ continuous main threads. Note that the implementation of $1$-D TCN follows the settings in \cite{bai2018empirical}. For testing the performance, we predict the $100$ randomly-selected main threads arrival time on both datasets, and the average MAE and RMSE in hours of all models are summarized in Table \ref{tab:main}, respectively. 
    
    As shown in Table \ref{tab:main}, TCN-related models consistently outperform other approaches in terms of MAE and RMSE of the prediction over both datasets. Explicitly, the ARIMA model only employs the time duration between events as the primary feature, which lacks the capability of learning the latent correlations in the time series prediction. Although RNN-based models and 1-D TCN can capture a more complex underlying relation between the arrival time of main threads, they cannot utilize the information from the associated replies, where may provide significant help in the prediction. Taking advantage of the grid structure, SocialGrid and its variants can achieve a more precise prediction result by considering the joint influence from other cascades without the assistance of additional features. 

    For SocialGrid and its variants, we also perform the ablation study to analyze the effect of different features in predicting the future main thread arrival time. By taking into account the relative temporal distance of each element towards the linked main thread, SocialGrid and SocialGrid-M can outperform the SocialGrid-S in predicting the main thread arrival time. Additionally, the prediction of the main thread arrival time appears to be unaffected by the information of $0_s$, since the prediction results between SocialGrid and SocialGrid-M are similar.
    
    \begin{figure}[tbp]
		\centerline{\includegraphics[width=0.49\textwidth]{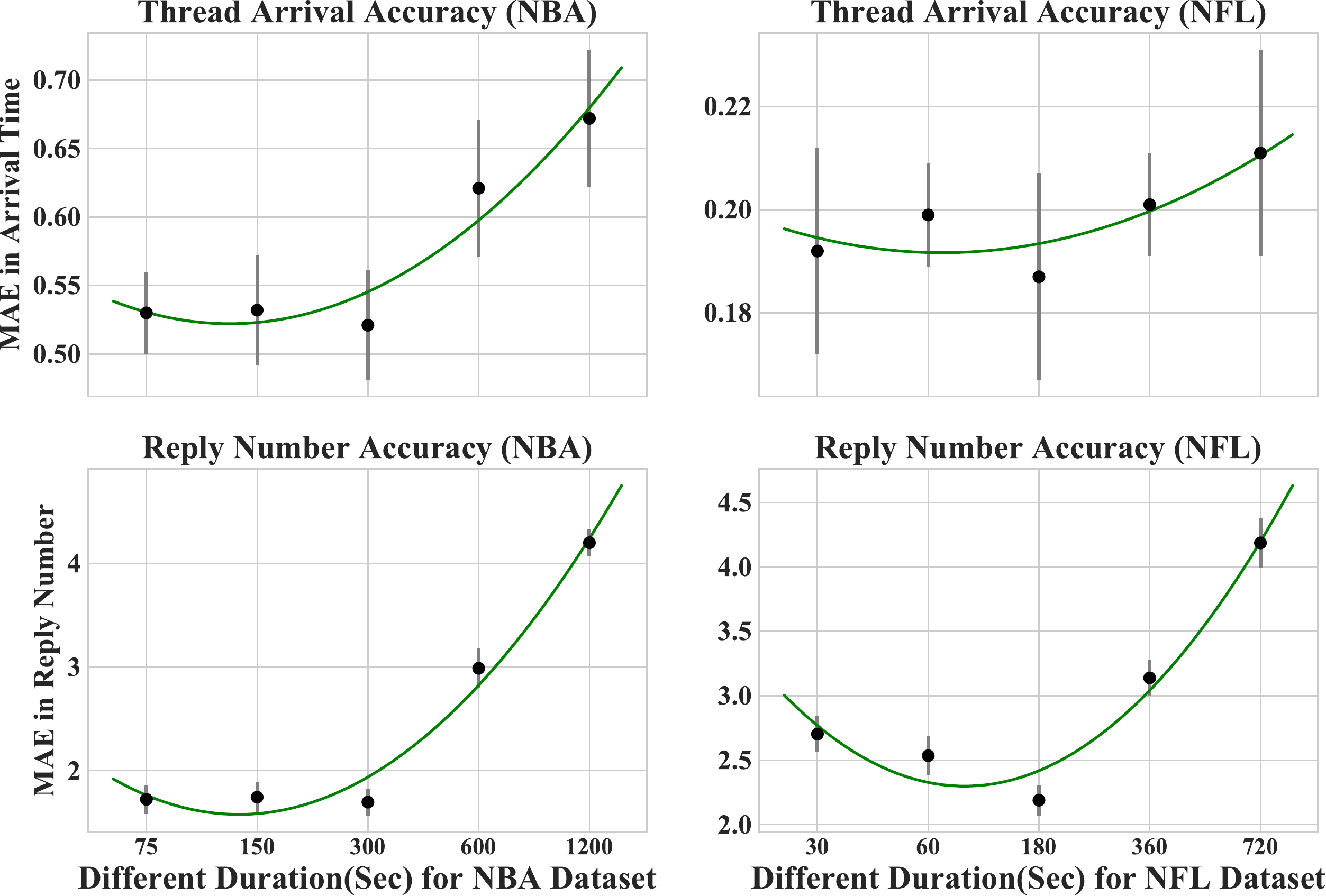}}
		\caption{The sensitivity of different time duration size $d$ in the prediction accuracy of both main thread arrival time and the number of reply.}
		\label{fig:5}
		\vspace{-2mm}
	\end{figure}
    
	\subsubsection{Prediction of Future Reply Number}
    We also try to predict the number of replies in the future time intervals. Note that this prediction often involves the cold start problem so that other approaches may not be able to handle appropriately. Therefore, we train other non-SocialGrid models on the arrival time information of $3,000$ replies from previous cascades. We assume that the evolution of replies under each cascade follows the same distribution. For testing, we randomly select $200$ thread-reply cascades out of $400$ cascades and make the prediction of the reply numbers in $20$ consecutive time intervals of the $200$ cascades. For non-SocialGrid methods, we adaptively simulate the arrival time of future replies and count the predicted number within one time interval. Note that $1$-D TCN is taken to be the variant of SocialGrid by changing the $K\times K$ filter size to $K\times 1$. The results are recorded in Table \ref{tab:reply}. 

    As shown in Table \ref{tab:reply}, TCN-based models still exhibit a lower prediction error compared with other approaches in both datasets. In particular, other non-SocialGrid approaches can only utilize univariate thread-reply cascade information as their training instances. However, each information cascade may have different evolution patterns. It is difficult for other non-SocialGrid models to learn all the reply evolutions in the same fitting function. Furthermore, compared with the $1$-D TCN, SocialGrid considers the mutual influence between information cascades, which leads to more robust prediction accuracy. For variants of SocialGrid, as the result of predicting the main threads, SocialGrid and SocialGrid-M outperform SocialGrid-S in the reply number prediction by employing the influence of relative time information. Additionally, as can be seen from the results between SocialGrid and SocialGrid-M, the information of $0_s$ in SocialGrid gives an improvement of the prediction accuracy.

	\begin{table*}[!t]
    \centering
    \resizebox{\textwidth}{!}{%
    \begin{tabular}{@{}c|c|c|c|c|c|c|c|c|c|c@{}}
    \toprule
    \multirow{2}{*}{Model} & \multicolumn{5}{c|}{NBA Dataset (MAE in Hrs)} & \multicolumn{5}{c}{NFL Dataset (MAE in Hrs)} \\ \cmidrule(l){2-11} 
    & 1 & 5 & 10 & 15 & 20 & 1 & 5 & 10 & 15 & 20 \\ \midrule
    \multicolumn{11}{c}{Temporal Point Process} \\ \midrule
    Hawkes & $0.885 \pm 0.82$ & $1.253 \pm 1.12$ & $1.479 \pm 1.31$ & $1.791 \pm 1.47$ & $2.492 \pm 1.68$ & $0.112 \pm 0.09$ & $0.151 \pm 0.12$ & $0.196 \pm 0.15$ & $0.217 \pm 0.19$ & $0.268 \pm 0.21$ \\ \midrule
    NesTPP & $0.792 \pm 0.68$ & $1.013 \pm 0.89$ & $1.232 \pm 0.84$ & $1.61 \pm 1.14$ & $1.972 \pm 1.37$ & $0.069 \pm 0.05$ & $0.091 \pm 0.08$ & $0.127 \pm 0.11$ & $0.161 \pm 0.17$ & $0.196 \pm 0.17$ \\ \midrule
    \multicolumn{11}{c}{Temporal Convolution Methods} \\ \midrule
    SocialGrid-S & $0.789 \pm 0.77$ & $1.114 \pm 0.97$ & $1.272 \pm 1.16$ & $1.598 \pm 1.31$ & $1.931 \pm 1.48$ & $0.062 \pm 0.06$ & $0.121 \pm 0.12$ & $0.132 \pm 0.13$ & $0.159 \pm 0.14$ & $0.217 \pm 0.19$ \\ \midrule
    \textbf{SocialGrid-M} & $\mathbf{0.707 \pm 0.71}$ & $\mathbf{0.981 \pm 0.84}$ & $\mathbf{1.158 \pm 1.06}$ & $\mathbf{1.437 \pm 1.14}$ & $\mathbf{1.812 \pm 1.27}$ & $\mathbf{0.046 \pm 0.07}$ & $\mathbf{0.089 \pm 0.11}$ & $\mathbf{0.106 \pm 0.11}$ & $\mathbf{0.139 \pm 0.18}$ & $\mathbf{0.179 \pm 0.19}$ \\ \midrule
    SocialGrid & $0.712 \pm 0.8$ & $0.985 \pm 0.97$ & $1.172 \pm 1.02$ & $1.455 \pm 1.01$ & $1.866 \pm 1.15$ & $0.05 \pm 0.07$ & $0.095 \pm 0.15$ & $0.118 \pm 0.19$ & $0.151 \pm 0.21$ & $0.188 \pm 0.24$ \\ \bottomrule
    \end{tabular}%
    }
    \caption{Adaptive Prediction of Future Main threads}
    \label{tab: ada_main}
    \end{table*}
    
    \begin{table*}[!t]
    \centering
    \resizebox{\textwidth}{!}{%
    \begin{tabular}{@{}c|c|c|c|c|c|c|c|c|c|c@{}}
    \toprule
    \multirow{2}{*}{Model} & \multicolumn{5}{c|}{NBA Dataset (MAE in Reply Number), $d=300$ sec} & \multicolumn{5}{c}{NFL Dataset (MAE in Reply Number), $d=180$ sec} \\ \cmidrule(l){2-11} 
    & $2d$ & $4d$ & $6d$ & $8d$ & $10d$ & $2d$ & $4d$ & $6d$ & $8d$ & $10d$ \\ \midrule
    \multicolumn{11}{c}{Temporal Point Process} \\ \midrule
    Hawkes & $22.378 \pm 6.32$ & $19.783 \pm 4.13$ & $15.251 \pm 4.12$ & $12.378 \pm 4.32$ & $9.427 \pm 2.38$ & $17.978 \pm 5.92$ & $14.519 \pm 3.79$ & $11.132 \pm 3.32$ & $9.763 \pm 3.12$ & $7.723 \pm 3.97$ \\ \midrule
    NesTPP & $15.562 \pm 5.62$ & $13.861 \pm 3.89$ & $10.865 \pm 3.12$ & $8.592 \pm 3.06$ & $7.681 \pm 2.79$ & $12.819 \pm 3.81$ & $10.032 \pm 3.65$ & $9.153 \pm 3.67$ & $7.598 \pm 2.61$ & $5.891 \pm 3.14$ \\ \midrule
    \multicolumn{11}{c}{Temporal Convolution Methods} \\ \midrule
    SocialGrid-S & $5.983 \pm 1.39$ & $5.241 \pm 1.13$ & $4.142 \pm 1.07$ & $3.571 \pm 0.98$ & $2.931 \pm 0.71$ & $3.124 \pm 0.69$ & $2.978 \pm 0.57$ & $2.756 \pm 0.48$ & $2.632 \pm 0.37$ & $2.495 \pm 0.31$ \\ \midrule
    SocialGrid-M & $5.342 \pm 1.07$ & $4.653 \pm 1.04$ & $3.771 \pm 0.96$ & $2.871 \pm 0.71$ & $2.161 \pm 0.47$ & $2.903 \pm 0.56$ & $2.582 \pm 0.39$ & $2.469 \pm 0.32$ & $2.318 \pm 0.36$ & $2.279 \pm 0.31$ \\ \midrule
    \textbf{SocialGrid} & $\mathbf{5.268 \pm 1.19}$ & $\mathbf{4.039 \pm 0.83}$ & $\mathbf{3.263 \pm 0.64}$ & $\mathbf{2.79 \pm 0.51}$ & $\mathbf{1.866 \pm 2.49}$ & $\mathbf{2.608 \pm 0.58}$ & $\mathbf{2.492 \pm 0.48}$ & $\mathbf{2.369 \pm 0.42}$ & $\mathbf{2.284 \pm 0.38}$ & $\mathbf{2.249 \pm 0.35}$ \\ \bottomrule
    \end{tabular}%
    }
    \caption{Adaptive Prediction of Reply Numbers}
    \vspace{-2mm}
    \label{tab: ada_reply}
    \end{table*}
    
    \subsubsection{Sensitivity to the Time Window}
	The size of $d$ can largely affect the prediction accuracy of both non-adaptive prediction applications, which is a fundamental setting in our experiment. Therefore, we analyze the prediction accuracy of both the thread arrival time and reply number prediction regarding the different values of $d$. Note that the value of the length $d$ is chosen based on the different time granularity of both datasets. For the prediction of future reply numbers, we choose to test the reply number in a different amount of time intervals based on the value of $d$.
    
    Fig. \ref{fig:5} shows the prediction accuracy of the main thread arrival time prediction and reply number prediction under different values of $d$. Over all four experiments on both datasets, the MAE curves fit a convex function, and there exists an optimal length $d$ that has the best prediction accuracy. For the NBA dataset, the optimal $d$ is $300$ seconds, since the MAE errors are the lowest in both predicting the future replies and main thread arrival time. Although the difference of prediction error is not evident among the time window $150$ and $300$ seconds, we stick with the $300$ seconds because of the faster data processing and training time. Similarly, the optimal time duration in the NFL dataset is $180$ seconds among the $30$, $60$, $360$, and $720$ seconds because of its lower prediction error among other settings. In the following experiments, we choose $d$ as $300$ and $180$ seconds in the NBA dataset and NFL dataset, respectively.
    
	\subsection{Adaptive Prediction of New Cascades}
	\subsubsection{Problem Description.}
	We describe our adaptive prediction procedure as follows: based on the proposed two models, given the full history $\mathcal{G}_{<i, \le j}$, we can know the arrival time of the $(j+1)$-th main thread by predicting the $O^{j}_{j+1}$. Then, we can start to simulate the future reply numbers that correspond to the $(j+1)$-th main thread in the following time intervals. By repeating this procedure, we can adaptively predict the evolution of thread-reply cascades in the future.
    
    The univariate time-series approaches (ARIMA and RNNs) cannot accomplish the adaptive prediction because of the cold start problem. Instead, we compare our model with the 1-D TPP since it models the base intensity of an event sequence (i.e., the possibility of observing a start event of the sequence). Therefore, we can use two separate temporal point processes to predict the arrival of new main threads and corresponding replies, respectively. Specifically, for the Hawkes process, two processes are trained separately with the arrival time information of $1,000$ main threads and $3,000$ replies similar to the settings in Sec. \ref{sec: non_adaptive}. The Hawkes process of the main thread first simulates the given number of main threads according to the provided history, and then the Hawkes process of replies simulates replies under each predicted main thread. For NesTPP, the model can automatically predict the future cascade evolution by the given history, we follow the same setting in the project \cite{chen2020}. Note that the multivariate TPP is only suitable for fix-dimensional sequence modeling, which is also not applicable in our case. Following the same setting in the non-adaptive prediction experiment, the evaluation of the adaptive prediction is also split into two parts: a) the accuracy of predicting the future main thread arrival time; b) the accuracy of predicting the future number of associated replies. The MAE errors of both prediction tasks on two datasets are summarized in Table \ref{tab: ada_main} and \ref{tab: ada_reply}, respectively.

    \subsubsection{Prediction Accuracy of Main Threads.}
	We randomly select $20$ $\mathcal{G}_{i, j}$ as the start point and simulate the arrival time of the following $20$ main threads with associated replies in the following $10$ blocks. The average MAE and standard deviation of the arrival time prediction are recorded in Table \ref{tab: ada_main}. As shown in the table, the average MAE of all models exhibits an increasing trend over time because of the error accumulation in the adaptive prediction. Although the distinction of prediction errors among each approach is not far behind, the proposed framework SocialGrid and its variants still have the overall lowest MAE comparing with the TPP methods in both datasets. Generally, SocialGrid and its variants are more capable of learning both short and long-range diffusion patterns by incorporating the influence of adjacent cascades. Furthermore, the adaptive prediction results among all the variants of SocialGrid exhibit the same trend as the non-adaptive main thread arrival time prediction. If we do not consider the relative information, the performance of SocialGrid-S is still the worst among the SocialGrid variants, and it is also worse than one of the TPP models (NesTPP). In addition, the SocialGrid-M still shows a generally lower error in adaptively predicting the main threads arrival time.

	\subsubsection{Prediction Accuracy of Reply Number.}
    Following the same procedure of simulating the main thread, we try to predict the number of replies under each forecasted main thread. On different datasets, we measure the number of predicted replies in the next $2d$, $4d$, $6d$, $8d$, and $10d$, where the $d$ is the length of the time interval in the Grid. According to Table \ref{tab: ada_reply}, SocialGrid and its variant models have an overall considerable improvement in adaptively predicting the reply numbers. Based on the information cascade \textit{dying-out} characteristic \cite{cheng2014can}, more replies generally arrive in the first few minutes after the posting time of the main thread, which explains the decreasing trend of MAE with the growth of the time intervals among all approaches. However, SocialGrid can better capture the history information from adjacent cascades to make a more accurate prediction. Similarly, by considering the information of the relative time and $0_s$, SocialGrid achieves the best prediction accuracy among other variants.

    \begin{figure}[!t]
		\centering
		\hspace{-3mm}\subfloat[Classification Correct Rate  (NBA)]{\label{fig:trend_nba}\includegraphics[width=0.24\textwidth]{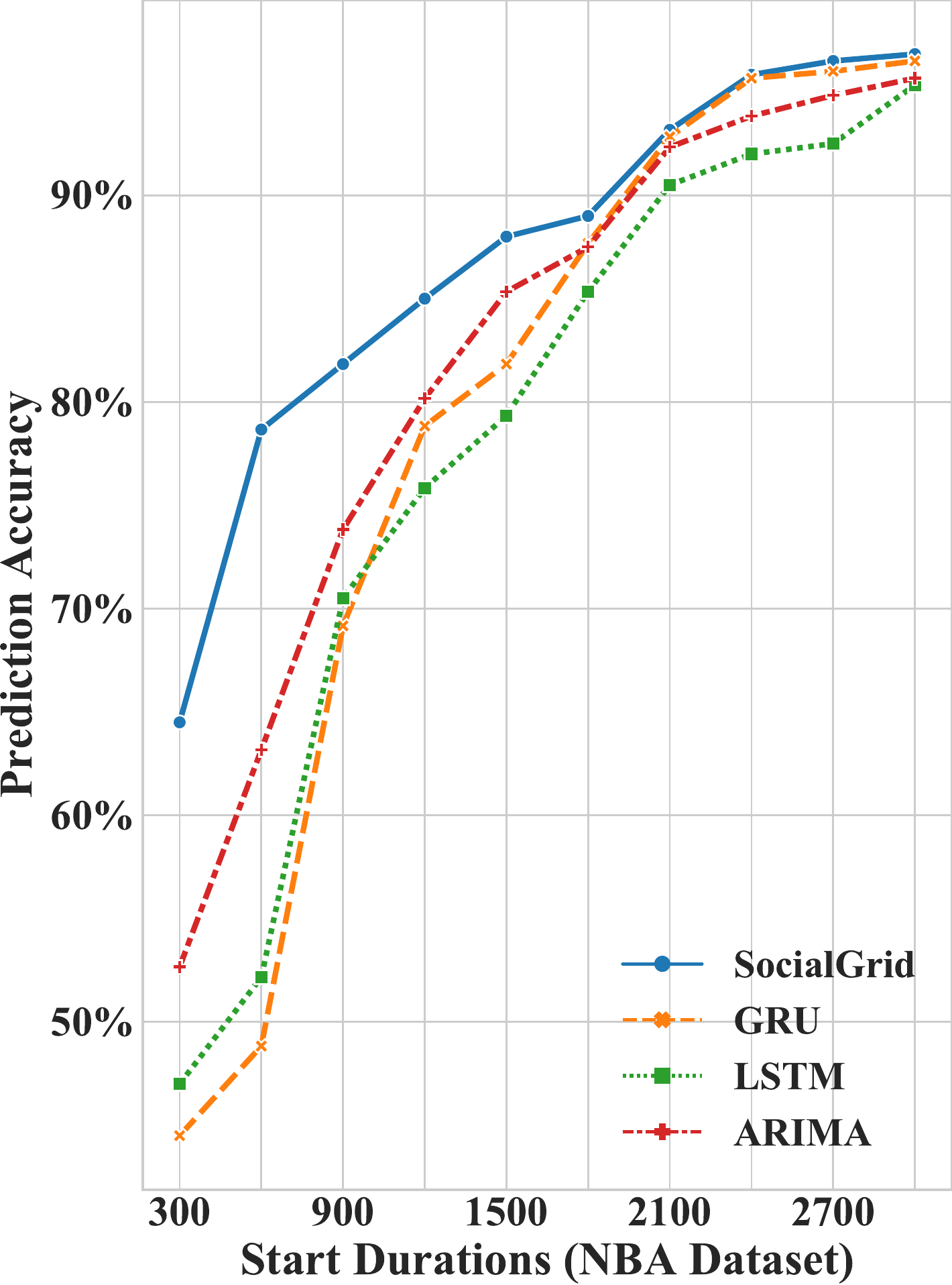}}
		\subfloat[Classification Correct Rate  (NFL)]{\label{fig:trend_nfl}\includegraphics[width=0.24\textwidth]{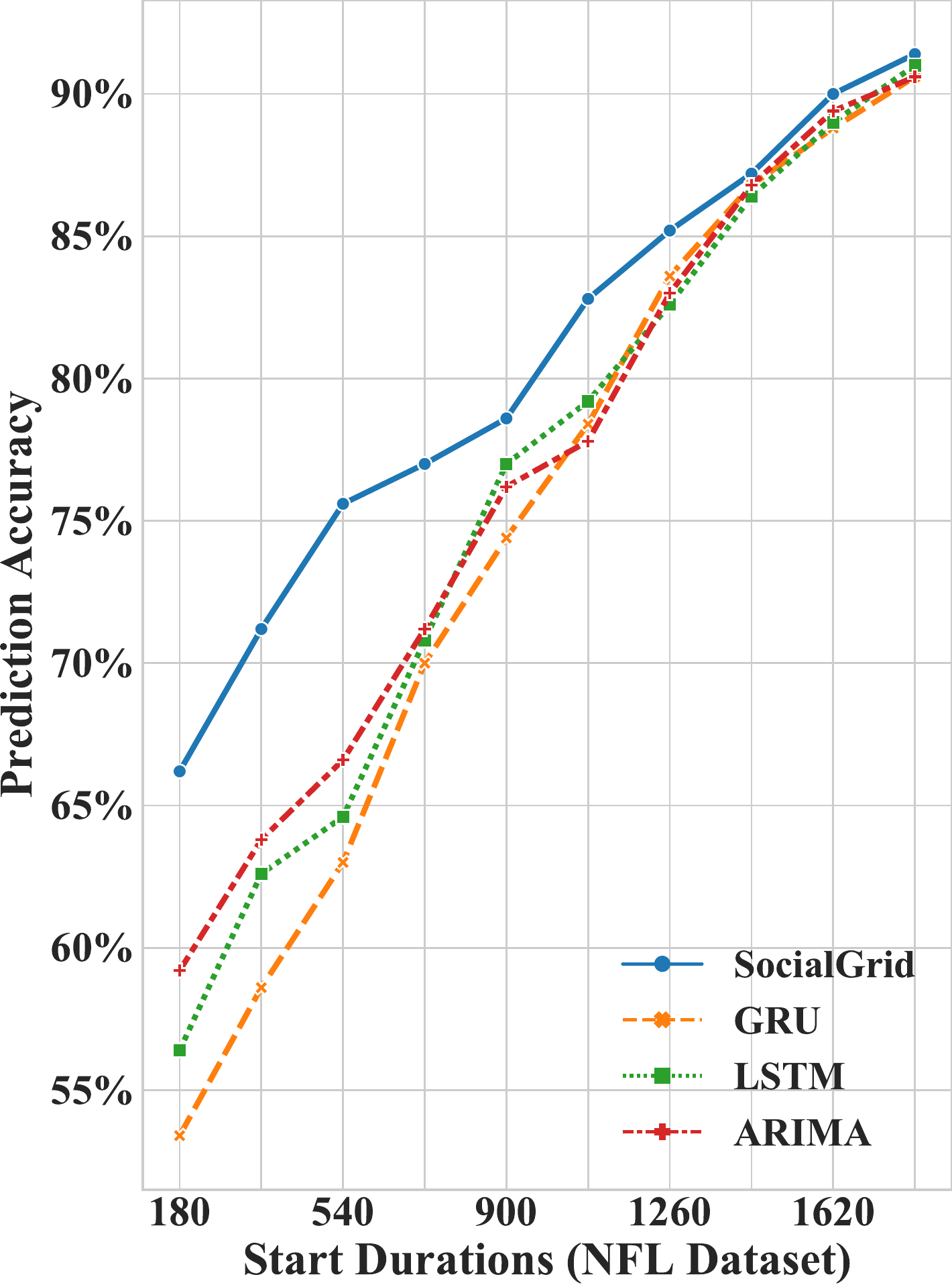}} 
		\hspace{3mm} 
		\caption{Correct Rate of Identifying Breakout Cascades: there are $60$ breakout cascades in the NBA dataset, and there are $55$ breakout cascades in the NFL dataset.}
		\label{fig:trend}
		\vspace{-2mm}
	\end{figure}

	\subsection{Application: Identifying Breakout Cascades}
	Can we accurately identify a breakout cascade before it receives the most attention? This task is significant in scenarios like rumor detection and product advertisement in ODFs. More specifically, we aim at forecasting if a thread-reply cascade would attract more than average replies by given the number of replies in the first a few minutes. We first find the average cascade size $\bar{L}$ among all cascades in our testing set. In our settings, the breakout cascades have the total number of replies that are more than twice the average cascade size $2\times \bar{L}$. We give all the approaches the prior information of reply numbers in the first few minutes with subject to the time duration of $d$ in SocialGrid. With the given prior knowledge, we employ the adaptive prediction approach to simulate the final volume of the cascade in the future, and the performance is taken to be the precision of successfully classifying a breakout cascade in the first $1d$, $2d$, ..., up to $10d$ seconds, which we call them the start durations. The classification results of both datasets are recorded in Fig \ref{fig:trend}.
    
    The essential of this task is to identify breakout cascades as early as possible. As we can see from both figures, SocialGrid can identify more breakout cascades in their early stage from both datasets. When the start durations are $300$ and $180$ seconds in both datasets, SocialGrid is able to identify around $65$\% of the breakout cascades without additional features in both datasets. With the size of the start duration increases, more information is revealed, which makes the breakout cascade more predictable. Eventually, the success rate of identifying breakout cascade of all approaches converge to nearly $100$\%, and SocialGrid still provides a steadily higher classification success rate among other methods. Overall, we prove that SocialGrid is capable of effectively detecting breakout cascades in ODFs.

	\section{Related Work}\label{sec: related}
	\textbf{Online Discussion Forum.} ODFs, as an essential branch in online social networks, have drawn substantial attention in recent years. Current researches on ODFs are ranging from user modeling \cite{goggins2016building, romero2013predicting} to content analysis and recommendation \cite{thomas2002learning, lan2018personalized}. However, the previous studies are conducted on the Massive Open Online Course platforms (i.e., a unique ODF for massive online education), studying the information diffusion problem in a more general ODF is needed. Moreover, the previous models incorporate auxiliary features to enhance the expressive power in various prediction tasks. However, as we mentioned earlier, the fundamental time-series prediction in information diffusion study of ODF requires long-range adaptively prediction, which is not applicable to utilize external mark information to enhance the prediction accuracy. In addition, very few works \cite{thomas2002learning, wang2011learning, medvedev2019modelling} have paid attention to the unique structure of the ODF in studying the information diffusion dynamics. The existing works still focus on the structural properties of thread-reply cascades and try to build a \textit{reply/discussion tree} for predicting the arrival time of future events, which still does not escape the confines of the traditional time-series analysis. In this work, we propose a grid-shaped way to represent an ODF, which can be learned by a more complex model.  

    \textbf{Time Series Prediction in Social Networks.} In the domain of information diffusion modeling and forecasting in the social network, extensive researches \cite{zhao2015seismic, bao2015modeling, kobayashi2016tideh} utilized the Gaussian-based point processes \cite{hawkes1971spectra} to predict the popularity of a post and the evolutionary dynamics of an information cascade through fitting the historical observations to a single intensity function. Such approaches, however, are unable to precisely predict the information dynamics in more complex real-world situations by the simple intensity function. Therefore, deep recurrent architectures \cite{bahdanau2014neural, luong2015effective} have been proposed to solve the shortcomings and achieved remarkable success in different social network sequence modeling tasks. Nevertheless, both temporal point process models and RNN models can only predict the fix-dimensional temporal sequences because of the \textit{cold-start} in dynamic prediction tasks \cite{schein2002methods}. In this work, to solve the \textit{cold-start} problem in the dynamic cascade prediction, we propose a novel convolution-based model that can predict the new thread-reply cascade by sliding the convolution filter on the proposed grid representation of an ODF.

    \textbf{Convolution Network for Time-series Forecasting.} Convolution networks are proven to have the ability to model sequence data back to the 90s \cite{sejnowski1987parallel, hinton1990connectionist}. Over time, CNNs have further been utilized in various sequence modeling tasks, including time-series classification \cite{cui2016multi, wang2017time}, machine translation \cite{gehring2016convolutional}, and audio recognition \cite{oord2016wavenet}, which show competitive results comparing with deep recurrent architectures. To date, inspired by the dilated convolution in \cite{oord2016wavenet}, TCN \cite{bai2018empirical} has been introduced as one of the strong competitors in various sequence modeling tasks. However, the traditional TCN model can only process univariate time-related sequences because of its one-dimension convolution architecture. The proposed framework SocialGrid is specially designed to process structured multivariate event sequences in the social discussion forum by transforming the ODF event space into a compressed grid representation.  
	
	\section{Conclusion and Future Work}\label{sec: con}
	\textbf{Conclusion.} In this paper, we propose a TCN enhanced deep learning framework - SocialGrid for modeling the information diffusion dynamics in ODFs. Through transforming the event space of an ODF into a grid-shaped representation, we have been able to utilize the proposed $2$-D TCN to capture underlying diffusion dynamics in the ODF and dynamically predict the evolution of a future thread-reply cascade. By only using time-series related features, SocialGrid brings flexibility to cascade prediction tasks as it only requires minimal and essential knowledge from information cascades, which can be extended to other social network platforms that have the similar thread-reply structure. Furthermore, extensive experiments and ablation studies show that our model outperforms different types of approaches on two real-world datasets, indicating its high potentials on exploring the temporal dynamics of thread-reply cascades with the help of the proposed grid structure. A practical application of identifying breakout cascades further proves that SocialGrid is not only theoretically sound but also has real-world uses.   
	
	\textbf{Future Works.} One of the promising future works is to incorporate more temporal features for both adaptive and non-adaptive predictions. Specifically, ODFs have active users from all over the world, and different time zones may affect the active period in particular subforums. Thus, the periodic information can be added to the SocialGrid as an exogeneous feature to consider the periodical influence in different time zones.

	
	\bibliographystyle{ACM-Reference-Format}
	\bibliography{cling}

\end{document}